          [2001/04/25 1.1 (PWD)]
\documentclass[a4paper,twocolumn]{esapub} 
\usepackage{natbib,graphicx}

\title{Resonant conversion of standing acoustic oscillations into Alfv{\'e}n waves in the 
$\beta{\sim}1$ region of the solar atmosphere}
\author{D. Kuridze$^1$}
\author{T.V. Zaqarashvili$^{1,2}$}
\author{B. Roberts$^3$}

\affil{\it 1. Georgian National Astrophysical Observatory (Abastumani
Astrophysical Observatory), Al. Kazbegi ave. 2a, 0160 Tbilisi,
Georgia, E-mail: dato.k@genao.org, temury@genao.org\\
\it 2. Departament de F\'{\i}sica,
Universitat de les Illes Balears, E-07122 Palma de Mallorca, Spain, E-mail: temury.zaqarashvili@uib.es\\
\it 3. School of Mathematics and Statistics, University of St. Andrews, St. Andrews, Fife, KY16
9SS, UK, E-mail: bernie@mcs.st-and.ac.uk}

\begin{document}


\maketitle

\begin{abstract}

We show that 5-minute acoustic oscillations may resonantly convert into 
Alfv{\'e}n waves in the $\beta{\sim}1$ region of the solar atmosphere. Considering the 
5-minute oscillations as pumping standing acoustic waves oscillating along 
unperturbed vertical magnetic field, we find on solving the ideal MHD equations that amplitudes of 
Alfv{\'e}n waves with twice the period and wavelength of acoustic waves exponentially 
grow in time when the sound and Alfv{\'e}n speeds are equal, 
i.e. $c_s \approx v_A$. 
The region of the solar atmosphere where this equality takes place we call a {\it swing layer}. 
The amplified Alfv{\'e}n waves may easily pass through the chromosphere and transition region carrying the energy of p-modes into the corona.

\end{abstract}

\section{Introduction}

It is generally considered that solar 5-minute photospheric acoustic oscillations cannot penetrate into the upper regions due to
the acoustic cutoff of stratified atmosphere. For the typical photospheric sound speed $c_s=7.5$ km/s the cutoff
frequency is 0.03 s$^{-1}$, which gives the cutoff period of 210 s \citep{rob1}. This means that the sound waves with 5 min ($\sim$ 300 s) period are evanescent. Acoustic oscillations cannot penetrate into the corona also due to
the sharp temperature gradient in the transition region. On the other hand, 3 and 5-minute intensity oscillations are intensively
observed in the corona by the space satellites SOHO (Solar and Heliospheric Observatory) and TRACE (Transition Region and Coronal
Explorer) \citep{dem}. Recently \citet{dep2} have discussed how photospheric oscillations can be channelled into 
the corona through inclined magnetic fields. Another solution of this controversy is the conversion of acoustic  
oscillations into another wave mode, which may pass through chromosphere/transition region.

Recent two dimensional numerical simulations \citep{ros,bog} outlined the importance of $\beta{\sim}1$ region in the solar atmosphere. They found the coupling of fast and slow magnetosonic waves at this particular region. 
Recent modelling of the plasma $\beta$ in the solar atmosphere (Gary 2001, see Fig. 3 of that paper) shows that 
$v_A \sim c_s$, i.e. $\beta \sim 1$ (actually $\beta \sim 1.2$ for $\gamma=5/3$), may takes place not only in lower chromosphere, but 
also at relatively low coronal heights (e.g., $\sim$ 1.2$R_0$ from the surface, where $R_0$ is the solar radius). Also latest observations \citep{mug}  suggest the possible transformation of compressible wave energy into incompressible waves at $\beta \approx 1$ region of the solar atmosphere. Thus this particular region may be of importance due to conversion of compressible wave energy into incompressible Alfv{\'e}n waves (or into MHD kink waves in thin photospheric magnetic tubes).  

The coupling of propagating sound and Alfv{\'e}n waves at $\beta{\sim}1$ has been proposed by Zaqarashvili and Roberts (2002,2005). They found that a sound wave is nonlinearly coupled to the Alfv{\'e}n wave with double the period and wavelength  when the sound and Alfv{\'e}n speeds are equal, i.e. $c_s \approx v_A$.

\citet{ulr} has reported observations of Alfv{\'e}n waves in the solar photosphere and lower chromosphere with substantial power at frequencies lower than the 5 minute oscillation. In the power spectrum of magnetic oscillations (Fig.3 in that paper) there is significant power at about 10 min. 

Here we show that standing acoustic waves oscillating along uniform magnetic field lines effectively generate 
Alfv{\'e}n waves with double their period and wavelength in the $\beta{\sim}1$ regions of the solar atmosphere.
The case of propagating waves is discussed in Zaqarashvili and Roberts (2002,2005).

\section[]{Statement of the problem and developments}

Consider fluid motions ${\bf u}$ in a magnetised medium (with zero
viscosity and infinite electrical conductivity), as described by
the ideal MHD equations:

\begin{equation}
{{{\partial \bf B}}\over {\partial t}} + ({\bf
u}{\cdot}{\nabla}){\bf B}= ({\bf B}{\cdot}{\nabla}){\bf u} - {\bf
B}{\nabla}{\cdot}{\bf u},
\end{equation}
\begin{equation}
{\rho}{{{\partial \bf u}}\over {\partial t}} + {\rho}({\bf
u}{\cdot}{\nabla}) {\bf u} = - {\bf {\nabla}}\left[p + {{B^2}\over
{8{\pi}}}\right ] + {{({{\bf B}{\cdot}{\nabla}}){\bf B}}\over
{4{\pi}}},
\end{equation}
\begin{equation}
{{{\partial {\rho}}}\over {\partial t}} + ({\bf
u}{\cdot}{\nabla}){\rho}
 + {\rho}{\nabla}{\cdot}{\bf u}=0.
\end{equation}
\begin{equation}
p=p_0\left ({{\rho}\over {\rho_0}}\right )^{\gamma},
\end{equation}
where $\rho$ is the medium density, $p$ is the pressure, ${\bf u}$
is the velocity, $\bf B$ is the magnetic field and $\gamma$ is the ratio of specific heats. For simplicity the stratification is neglected here, but we plan to take it into account in future.

Cartesian coordinate system is adopted with the $z$ axis directed vertically upwards from the solar surface. 
Spatially inhomogeneous (along the $x$ axis) magnetic field is directed along the $z$ axis (see Fig.1), i.e. 
\begin{equation}B_0=(0,0,B_z(x)).
\end{equation}

Plasma pressure and density also are assumed to have $x$ dependence, so they are expressed as $p_0(x)$ and $\rho_0(x)$ respectively.
The magnetic field and pressure satisfy the transverse pressure balance condition
\begin{equation}
p_0(x) + {{B^2_z(x)}\over {8\pi}}=const.
\end{equation}  

Plasma $\beta$ is defined as 
\begin{equation}
\beta = {{8\pi p_0(x)}\over {B^2_z(x)}}= {{2c^2_s}/{\gamma v^2_A(x)}},
\end{equation}
where $c_s=\sqrt{\gamma p_0/\rho_0}$ and $v_A(x)=B_z/\sqrt{4 \pi \rho_0}$ are the sound and Alfv{\'e}n
speeds respectively. Note that we suggest the temperature to be homogeneous so the sound speed does no depend on the $x$ coordinate.      
       
We consider wave propagation along the $z$ axes (thus along the magnetic field) and 
wave polarisation in $yz$ plane. Then only sound and Alfv{\'e}n waves arise. The velocity component of sound wave is polarised along the 
$z$ axis and the velocity component of the Alfv{\'e}n wave is polarised along the y-axis. In this case equations (1)-(3) take form

\begin{equation}
{{{\partial b_y}}\over {\partial t}} + u_z{{{\partial b_y}}\over
{\partial z}} = - b_y{{{\partial u_z}}\over {\partial z}} +
B_z(x){{{\partial u_y}}\over {\partial z}},
\end{equation}
\begin{equation}
{\rho}{{{\partial u_y}}\over {\partial t}} + {\rho}u_z{{{\partial
{u_y}}}\over {\partial z}} = {{B_z(x)}\over {4\pi}}{{\partial
b_y}\over {\partial z}},
\end{equation}
\begin{equation}
{{{\partial {\rho}}}\over {\partial t}}= - {\rho}{{{\partial
{u_z}}}\over {\partial z}} - u_z{{{\partial {\rho}}}\over
{\partial z}},
\end{equation}
\begin{equation}
{\rho}{{{\partial u_z}}\over {\partial t}} + {\rho}u_z{{{\partial
{u_z}}}\over {\partial z}} = - {{{\partial p}}\over {\partial z}} -
{{\partial}\over {\partial z}}{{b^2_y}\over {8\pi}},
\end{equation}
\begin{equation}
{{{\partial p}}\over {\partial t}} = - {\gamma}p{{{\partial
{u_z}}}\over {\partial z}} - u_z{{{\partial p}}\over {\partial
z}},
\end{equation}

where $p=p_0 + p_1$ and $\rho={\rho}_0 + \rho_1$ denote the total (unperturbed plus perturbed) pressure and density, $u_y$
and $u_z$ are the velocity perturbations (of the Alfv{\'e}n and
sound waves, respectively), and $b_y$ is the perturbation in the
magnetic field. Note, that in these equations the $x$ coordinate stands as a parameter.
 
\begin{figure}
\centering
\includegraphics[width=1\linewidth]{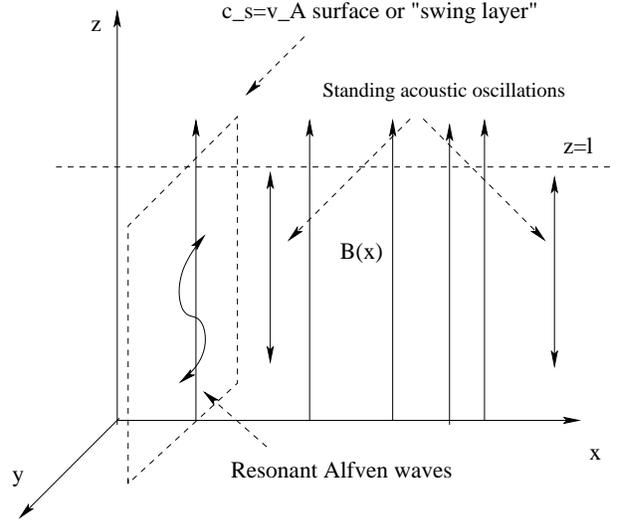}
\caption{ Schematic picture of the proposed process. 
The $z$ coordinate is directed vertically upwards, while the $x$ and $y$ 
axes are parallel to the solar surface.
The magnetic field and density are inhomogeneous along the $x$ coordinate. 
Acoustic waves propagate up and down, i.e. along the $z$ axis. 
Alfv{\'e}n wave velocity oscillates along the $y$ axis.
Standing acoustic waves are resonantly converted into Alfv{\'e}n waves near the region where $v_A \sim c_s$ called as {\it swing layer}.\label{fig:single}}
\end{figure}

Acoustic waves oscillate along the $z$ axis, so that the velocity component has nodes at points 
$z=0$ and $z=l$, i.e. $u_z=0$ at $z=0, l$, and thus we take 
\begin{equation}
u_z=v(t)\sin(k_sz),
\end{equation}
\begin{equation}
\rho_1 = \rho_1(t)\cos(k_sz),
\end{equation}
where $k_s$ is the wavenumber of sound wave such that 
$$
k_sl={{2\pi l}\over {\lambda_s}}= n\pi,
$$
so 
$$
{l\over {\lambda_s}}=n/2,
$$
where $n=1,2...$.   

We express the Alfv{\'e}n wave components as
\begin{equation}
b_y=b(t)\cos(k_Az),
\end{equation}
\begin{equation}
u_y=u(t)\sin(k_Az),
\end{equation}
where $k_A$ is the wavenumber of the Alfv{\'e}n waves.

Substitution of expressions (13)-(16) into equations (8)-(12) and averaging with $z$ over the distance $(0,l)$ leads to the cancelling
of all nonlinear terms, which means that waves do not interact. However in particular case, when wave numbers $k_s$ and $k_A$ satisfy the conditions
\begin{equation}
k_s = 2k_A,
\end{equation}
equations (8)-(12) take the form:
\begin{equation}
{{{\partial b}}\over {\partial t}} = k_AB_0u(t) - {k_A\over 2}v(t)b(t),
\end{equation}
\begin{equation}
{{{\partial u}}\over {\partial t}} = - {{k_AB_0}\over {4\pi\rho_0}}{b(t)\over {1 - {{\rho_1(t)}\over {2\rho_0}}}} - {k_A\over 2}{{u(t)v(t)}\over {1 - {{\rho_1(t)}\over {2\rho_0}}}},
\end{equation}
\begin{equation}
{{{\partial v}}\over {\partial t}} = {{k_sc^2_s}\over
{\rho_0}}\rho_1(t) + {{k_A}\over {8\pi\rho_0}}b^2(t),
\end{equation}
\begin{equation}
{{{\partial \rho_1}}\over {\partial t}} = - \rho_0k_sv(t),
\end{equation}
which means that the waves may interact as the nonlinear terms remain.

Substitution of $u$ from equation (18) into equation (19) and
neglecting of all third order terms leads to the second order
differential equation
$$
{{{\partial^2 b}}\over {\partial t^2}} + k_Av{{{\partial b}}\over
{\partial t}} + \left [k^2_Av^2_A + {{k^2_Av^2_A}\over
{2\rho_0}}\rho_1 + {k_A\over 2}{{\partial v}\over {\partial t}}
\right ]b=
$$
\begin{equation}
 = 0.
\end{equation}
This equation describes the time evolution of Alfv{\'e}n wave spatial Fourier harmonics expressed by (15)-(16) forced by standing acoustic waves.  

In this equation, the first derivative with time can be avoided by substitution of function 
$$
b={\tilde b}(t)e^{-{k_A\over 2}{\int
{vdt}}},
$$
which after dropping third order terms leads to
\begin{equation}
{{{\partial^2 {\tilde b}}}\over {\partial t^2}} + \left [k^2_Av^2_A
+ {{k^2_Av^2_A}\over {2\rho_0}}\rho_1 \right ]{\tilde b}= 0.
\end{equation}

Equation (23) reflects the fact that the Alfv{\'e}n speed is modified due to the density variation of standing acoustic wave. The similar equation 
for $\beta \gg 1$ was obtained by \citet{zaq0}. It is seen from this equation that the particular time dependence of density perturbation determines the type of equation and consequently its solutions. If we consider the initial amplitude of Alfv{\'e}n waves smaller than the amplitude of acoustic waves, then the term with $b^2$ in equation (20) can be neglected. This means that the backreaction of Alfv{\'e}n waves due to the ponderomotive force is small. Then the solution of equations (20)-(21) is just harmonic function of time
\begin{equation}
\rho_1 = \alpha \rho_0 \cos(\omega_s t),
\end{equation}
where $\omega_s$ is the frequency of standing acoustic wave and $\alpha \ll 1$ is the relative amplitude. Here we consider the small amplitude acoustic waves $\alpha \ll 1$, so the nonlinear steepening due to the generation of higher harmonics is negligible. Then the substitution of expression (24) into equation (23) leads to the Mathieu equation
\begin{equation}
{{{\partial^2 {\tilde b}}}\over {\partial t^2}} + k^2_Av^2_A\left [1
+ {{\alpha}\over 2}\cos(\omega_s t) \right ]{\tilde b}= 0.
\end{equation}

The solution of this equation with frequency ${\omega_s}/2$ has an exponentially growing character, 
thus the main resonant solution occurs when \citep{zaq1,she}
\begin{equation}{\omega_A}= v_Ak_A ={{\omega_s}\over 2},\end{equation}
where $\omega_A$ is the frequency of Alfv{\'e}n waves. Since $k_A=k_s/2$, resonance takes place when
\begin{equation}v_A =c_s.\end{equation}

Since the Alfv{\'e}n speed $v_A(x)=B_z(x)/\sqrt{4 \pi \rho_0(x)}$ is a function of the $x$ coordinate, 
then this relation is satisfied at a particular location along the $x$ axis (see Fig. 1). 
Therefore near this region the acoustic oscillations will be resonantly transformed into Alfv{\'e}n waves. 
We call this region the {\it swing layer}, by analogy 
with mechanical swing interactions (see a similar consideration in \citet{she}).       

Under condition (26) the solution of equation (25) is 
\begin{equation}
{\tilde b}(t)={\tilde b}_0e^{{{\left
|{\alpha}{\omega_s}\right |}\over 16}t}\left
[{\cos}{{\omega_s}\over 2}t {\mp} {\sin}{{\omega_s}\over 2}t
\right ],
\end{equation}
where ${\tilde b}_0={\tilde b}(0)$ and the phase sign depends on
${\alpha}$; it is $+$ for negative ${\alpha}$
and $-$ for positive ${\alpha}$. 

Note that the solution
has a resonant character within the frequency interval
\begin{equation} {\left |{\omega_A} -
{{\omega_s}\over 2} \right |}<{\left |{{\alpha \omega_s}\over 8} \right |}.
\end{equation}

This expression can be rewritten as
\begin{equation} {\left |{{v_A}\over c_s} - 1 \right |}<{\left |{{\alpha }\over 4} \right |}.
\end{equation}
Thus the thickness of the {\it swing layer} depends on the acoustic wave amplitude!  

Therefore the acoustic oscillations are converted into Alfv{\'e}n waves not only at the surface $v_A =c_s$ but also near 
that region, namely at 
\begin{equation}v_A = c_s \left (1 \pm {{\alpha }\over 4}\right ).\end{equation}
Thus the resonant layer can be significantly wider for stronger amplitude acoustic oscillations.

\begin{figure}
\centering
\includegraphics[width=1\linewidth]{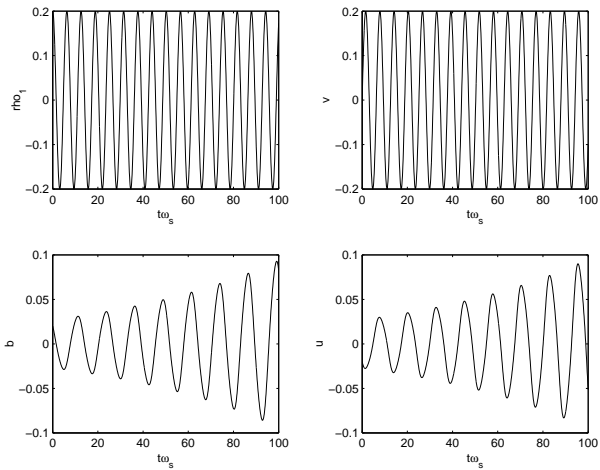}
\caption{Numerical simulations of wave conversion. 
Upper panels show the density and velocity components of standing acoustic oscillations, 
while lower panels show the magnetic field and velocity components of Alfv{\'e}n waves. 
Alfv{\'e}n waves with twice the period of acoustic oscillations grow in time. \label{fig:single}}
\end{figure}

Numerical solution of equations (18)-(21) (here the backreaction of Alfv{\'e}n waves is again neglected) is presented on Fig.2. 
The amplification of Alfv{\'e}n waves with double the period of acoustic oscillations is clearly seen.

\section{Conclusions}

We suggest that 3 and 5-minute acoustic oscillations in 
photosphere/chromosphere can be resonantly converted into Alfv{\'e}n waves, or possibly into MHD kink waves in thin
photospheric magnetic tubes, this process acting in the region of
the solar atmosphere where $v_A{\approx}v_s$. Generated
transversal waves may then propagate through the transition
region into the corona, where they can deposit their energy back
into density perturbations. The process can thus be of importance 
in coronal heating.

\section{Acknowledgements}

D. Kuridze and T. Zaqarashvili acknowledge the financial support from conference organisers.

\end{document}